# Shuttlecock velocity decay after smash and slice shots in badminton

Eric Collet[1,2,*]

[1] Univ Rennes, CNRS, IPR (Institut de Physique de Rennes) – UMR 6251; 35000 Rennes, France.
[2] Institut Universitaire de France (IUF); 75231 Paris, France

*Author to whom any correspondence should be addressed.

E-mail: eric.collet@univ-rennes.fr

## Abstract

This paper reports on the decay in shuttlecock velocity after smash and slice shots performed by elite and international players, based on the analysis of videos captured on a badminton court with high-speed cameras. The trajectories of feather shuttlecocks show an exponential decay in velocity and an exponential increase in time-of-flight with distance, in accordance with the equation of motion of a shuttlecock subjected to high drag. The initial speed, which can exceed 500 km/h, is reduced by half every $\approx 3.35$ m, depending on the physical parameters of the shuttlecock characterized by a speed index. The videos show that plastic shuttlecocks, which are more resistant than feather shuttlecocks, deform at high speeds. The resulting decay of their drag coefficient with increasing speed makes them unsuitable for high-level play. The study also shows that the spin induced by the slice shots of left-handed players or the reverse slice shots of right-handed players slows the shuttlecock down.

## 1. Introduction

The smash and slice in badminton are tactical shots, usually played from the back of the court, that often score a point. The smash, hit with power and speed, is also a source of fantasy making badminton the fastest sport in the world: P. Tan set the women's record at 438 km/h, while S. Rankireddy set the men's record at 565 km/h. Numerous studies have focused on the particular trajectories of shuttlecocks, which result from the strong aerodynamic drag experienced by these light ($\approx 5$ g) and elongated ($\approx 7$ cm) particles (1-8). This drag is essential for keeping the shuttlecock within the confines of the court, as it strongly decreases the velocity of the shuttlecock (9-16): an initial velocity of $\simeq 240$ km/h (67 m/s) can decay within $\simeq 0.6$ s to $\simeq 25$ km/h (6.9 m/s).

The speed of the shuttlecock depends on the energy transfer between the player, the racket and the shuttlecock, as underlined by diverse bio-mechanical studies (17-22).

This article presents a quantitative analysis of the trajectories and velocity decay of feather and plastic shuttlecocks after smash shots performed by international and elite players. The effect of the shuttlecock's physical parameters, in particular its mass, is also discussed. These can be adapted on demand to ensure similar playing conditions depending on atmospheric conditions on the planet. The effect of slice and reverse-slice shots on velocity decay is also analysed, for both right- and left-handed players. Some tactical conclusions are discussed, based on experimental and theoretical results. Considering the expected readership, the choice is made to use both the popular velocity unit (km/h) and the metric unit (m/s).

**Table 1.** Terminal velocity $V_T$ measured for different references of shuttlecocks with different speed index, mass $m$ and radius $R$.

| Speed index | AS20 76/V2 | AS20 77/V3 | AS20 78/V4 | Mavis 300 |
|---|---|---|---|---|
| $m$ (g) | 5.10(2) | 5.25(2) | 5.45(2) | 5.30(5) |
| $R$ (cm) | 3.35(5) | 3.35(5) | 3.35(5) | 3.30(2) |
| $V_T$ (m/s) | 6.19(2) | 6.34(2) | 6.48(2) | 6.05(2) |

## 2. Method

A Phantom Miro 3a10 high-speed camera was used on the badminton court to capture high frame rate videos (3000 fps) of shuttlecock trajectories. The resolution of the videos was maximized to 1280×800 pixels with an exposure time of 333 µs for each frame. A Canon camera was also used, with a capture frame rate set to 500 fps (1920×1080 pixels). The videos captured the shots performed on court by elite national and international players, during the French international badminton championships 2025 (IFB) in Rennes and during a session dedicated to the present study. Tracker (23, 24) a free Java video analysis tool, was used to track shuttlecock trajectories and extract speed or time-of-flight curves.

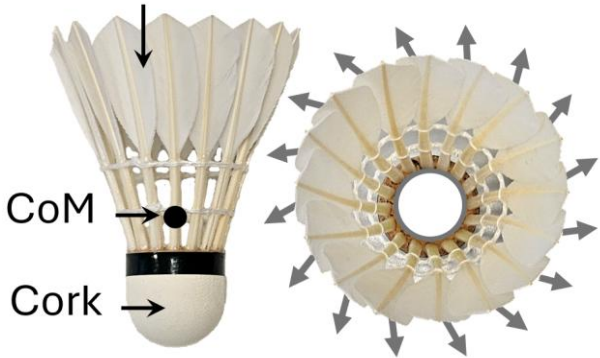

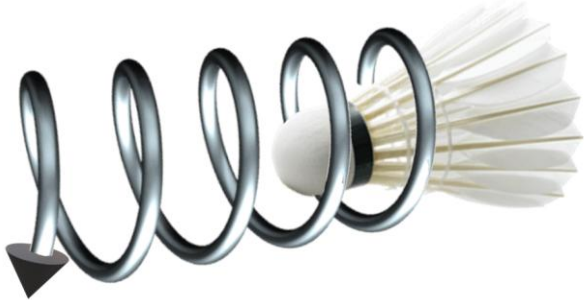

**Figure 1.** Structure of a feathered shuttlecock. The air flow on the 16 feathers (arrows) induces a natural counter-clockwise spinning.

## 3. Drag force and shuttlecock's speed index

Figure 1 shows the structure of feather shuttlecocks, composed of an array of diverging stems, the ends of which are at the convergent end of the skirt, joined together in an end ring. Shuttlecocks do not exhibit mirror symmetry and are chiral bodies (25). Due to the way the 16 feathers are placed into the cork, the airflow on the feathers is responsible for the natural counter-clockwise spinning of these projectiles as they propagate through the air. The centre of mass (CoM) is on the shuttlecock symmetry axis, close to the cork.

The aerodynamics of shuttlecocks was studied in several papers, based on wind tunnel measurements, video analysis or simulations (1-8, 13, 15, 26, 27). The equation of motion of a shuttlecock was given in many papers and here we use:

$$m\frac{d\vec{V}}{dt} = m\vec{g} - \vec{F_D} \quad (1)$$

where $\vec{g}$ is gravity, m is the mass of the shuttlecock, $\vec{V}$ its velocity and $\vec{F_D}$ the drag force. This one depends on the air density $\rho$, the cross-section of the shuttlecock ($\pi R^2$) with radius $R$ and the drag coefficient $C_D$. The equation reaches:

$$m\frac{d\vec{V}}{dt} = m\vec{g} - \frac{C_D}{2}\pi R^2 \rho V \vec{V} \quad (2)$$

Peastrel *et al* studied the terminal velocity of a shuttlecock in vertical fall (28), where gravity accelerates the shuttlecock, which increases the drag force until weight and drag force balance each other. Then, the projectile reaches its terminal velocity $\vec{V_T}$ in free fall. The velocity is then constant ($\frac{d\vec{V}}{dt} = 0$), with $\vec{V_T}$ anticolinear to $\vec{g}$ (2), which gives:

$$V_T = \sqrt{\frac{2mg}{C_D \rho \pi R^2}} \quad (3)$$

The typical values for $m$=5.25 g, $\rho$=1.2 kg m$^{-3}$, $\pi R^2$=35 cm$^2$ and $C_D$=0.6, give a terminal velocity $V_T \approx$24 km/h (or 6.5 m/s).

On earth, $\rho$ changes with various parameters, such as atmospheric pressure, altitude, temperature or humidity. Higher air density causes more drag, which strongly affects $V_T$ and the flying distance of the shuttlecock. In order to ensure similar playing conditions across different locations around the world, the parameters of the shuttlecocks (mass, radius and/or $C_D$) are adapted to atmospheric conditions. The badminton world federation (BWF) uses shuttlecocks with different "speed index" ranging from 75 (or V1) for slower shuttlecocks to 79 (or V5) for faster shuttlecocks. Table 1 shows the evolution of $V_T$ measured for a plastic shuttlecock (Mavis 300) and feather shuttlecocks Yonex AS20 with different speed index (76/V2 to 78/V4). The AS20 shuttlecocks used here had similar radius. Their increasing mass, correlates with the increasing speed index and $V_T$.

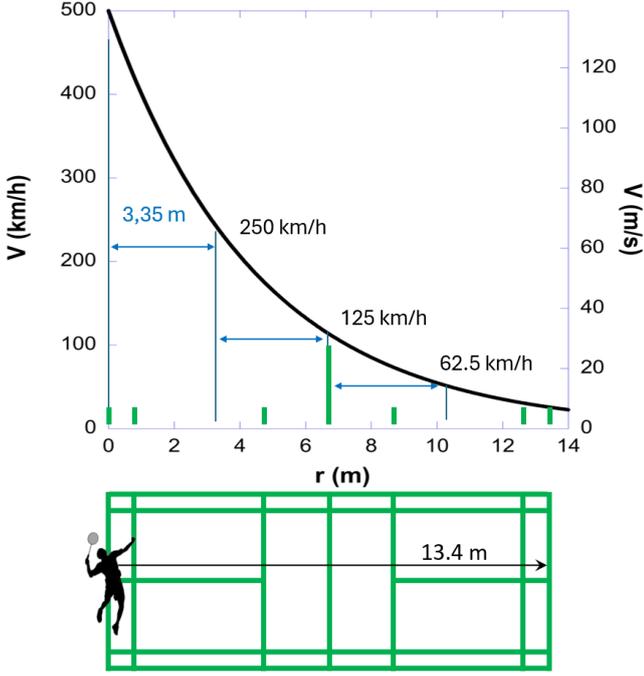

**Figure 2.** Theoretical exponential decrease of the velocity of the badminton shuttlecock along the court, for an initial speed of 500 km/h (≈139 m/s) and a velocity halving distance $L_{1/2}$≈3.35 m.

## 4. Exponentioal decay of the schuttlecock velocity

### 4.1 Theoretical decay of the shuttlecock velocity

Lets consider the velocity of a shuttlecock reaching 500 km/h, as often reported in competition. Since the drag force $F_D$ at $V_T$≈25 km/h equals weight in free fall, and since $F_D \propto V^2$, the drag force is about 400 times larger than weight after such a smash stroke. Weight is then negligeable at high speed. The equation of motion is then simplified, with the drag force anticolinear to the velocity:

$$m\frac{dV}{dt} = -\frac{1}{2}\rho\pi R^2 C_D V^2 \quad (4)$$

By introducing the infenitesimal travel $dr=Vdt$ of the shuttlecock during $dt$, (4) can be rewritten :

$$\frac{dV}{V} = -\frac{1}{2m}\rho\pi R^2 C_D dr \quad (5)$$

and integrating (5) gives :

$$V(r) = V_0 \exp\left(-\frac{\rho\pi R^2 C_D}{2m}r\right) = V_0 \exp\left(-\frac{r}{L}\right) \quad (6)$$

Equation 6 describes the exponential decay in shuttlecock velocity with the distance $r$, for an intial velocity $V_0$. $L = \frac{2m}{\rho\pi R^2 C_D}$ is called the aerodynamic length and a typical value is $L$≈4.5 m (2). For general readers, it is convenient to use $L_{1/2} = \frac{2}{e}L \approx 3.35\,m$, which is the characteristic velocity halving distance. Figure 2 shows the theoretical velocity decay from equation (6), for an initial speed $V_0$=500 km/h (139 m/s) from the backline of the badminton court. The length of the court is 13.4 m ≈ $4L_{1/2}$, which means that the shuttlecock velocity is divided by 4 (125 km/h) at the net and by16 (≈31 km/h) at the opposite backline of the court, which approaches $V_T$. Then the weight can't be neglected anymore to accurately describe the trajectory. The range of validity of equation (6) to describe the trajectory is therefore questionable.

Figure 3a shows the trajectory of the center of mass of shuttlecock Yonex AS50 77/V3 after a smash shot (performed by T. Jr. Popov during the IFB 2025).

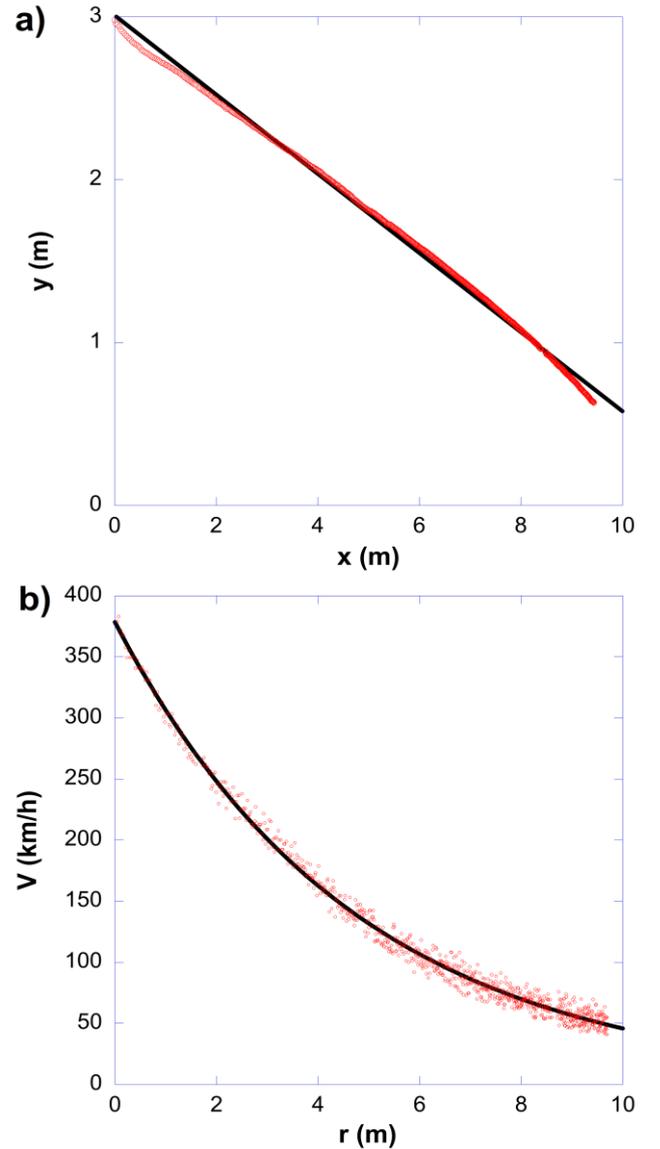

**Figure 3.** a) Shuttlecock trajectory: x is the coordinate along the court, y is the altitude. b) Velocity decay with the distance $r$ from the smash (dots). The black line is the exponential fit.

The (x,y) trajectory is extracted from the video analysis, where x is the horizontal distance along the long court axis and y is altitude of the shuttle. The shot is performed in the downward direction to hit the court. The trajectory is fairly close to a linear trajectory (black line). The y coordinate is affected by gravity during the time-of-flight of 300 ms. However, after 9.5 m and a time of flight of 300 ms, the y coordinate deviates only by 10 cm from the linear trajectory. This is less than the displacement of a free falling object in gravity during a time-of-flight of 300 ms, $\frac{1}{2}gt^2 = 45\ cm$. Indeed, the shuttlecock propagating in the air is not a free falling object, due to the important drag force.

Figure 3b shows the measured velocity decay $V(r)$ of the shuttlecock with the distance $r$ from the point of impact with the racket. The fit of $V(r)$ with the exponential law (black curve) given by equation (6) is in fairly good agreement with the experimental data and gives an initial velocity $V_0$=379(1) km/h and aerodynamic length $L$=4.73(2) m. $L$ is close to the theoretical value, but it can strongly depend on the shuttlecock characteristic and the potential damage it sustains during gameplay, which alters the drag force.

Overall, it is reasonable to consider that equation (6) is relatively accurate to describe the shuttlecock trajectory after a smash with a high initial velocity and up to the first 10 m.

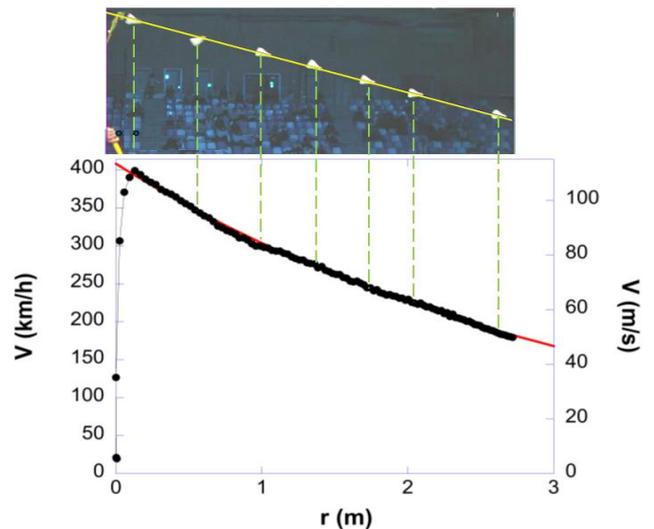

**Figure 5.** Deviation of the velocity of the CoM from the exponential law (red fit). The deceleration is larger when the shuttle axis is not aligned along the trajectory (yellow line).

### 4.2 The velocity of what? Cork vs centre of mass (CoM).

Given that a badminton shuttlecock is not a point object, one might wonder which velocity should be measured. Should we take into account the speed of the CoM or that of the cork? Figure 4 shows the evolution of both velocities after a smash stroke, performed by A.S. Rasmussen during the IFB 2025 with a Yonex AS50 77/V3 (video 1). The fit of the velocity of the CoM with an exponential law (blue curve) fairly agrees with the data and gives an initial velocity $V_0$=396(1) km/h and an aerodynamic length $L$=4.17(2) m. The speed of the cork strongly deviates from the exponential law during the first meter, due to the change in orientation of the shuttle from almost vertical to horizontal: the initial speed of the cork is much higher (451(1) km/h, figure 4) than the one of the CoM (396(1) km/h). Finaly, both velocities converge (within the experimental error bar) once the axis of the shuttle is aligned along the trajectory. Given that the BWF rules consider the point of impact of the cork on the court to award points, one might wonder whether it would not be more relevant to take into account its speed rather than that of the centre of mass. Herafter $V_0$ refers to the initial speed of the CoM.

Figure 5 shows a closer inspection of the velocity of the CoM of the shuttle, monitored over the first 2.5 m (shot by C.-L. Wang during the IFB 2025, video 2). The data reveal damped oscillation of the CoM velocity, compared to the exponential velocity decay fit (red curve). The correlation between the chronograph and $V(r)$ highlights that the decelatation is higher when the shuttlecock axis is not aligned along the trajectory. Cohen *et al* explained this flipping process of the shuttle (2) as its axis aligns axis along the velocity direction. Figure 4 shows that the oscillating componant is damped within ≈1 m for such high speed.

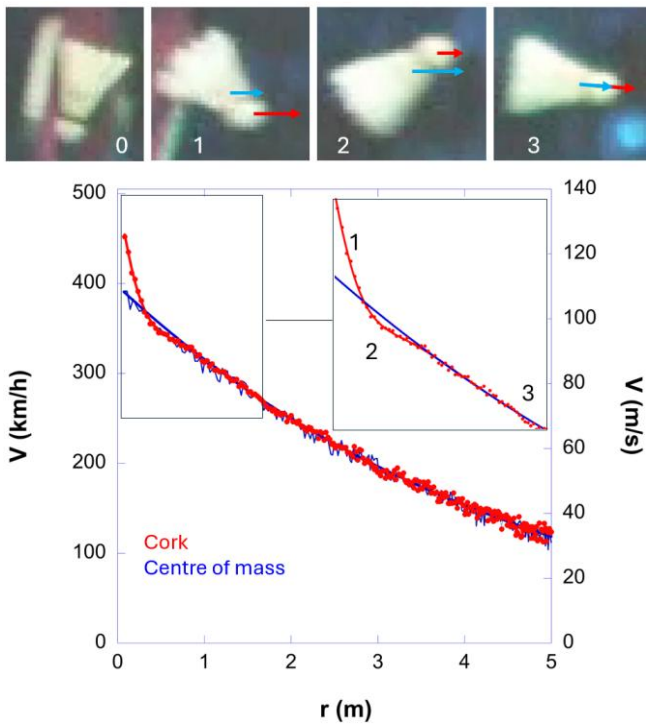

**Figure 4.** Velocities (blue and red arrows) of the CoM of the shuttlecock and of the cork, altered by the orientation from almost vertical (0) to almost horizontal after the smash. The cork is faster (1) and slower (2) than the CoM and both velocities are equal once oriented along the axis of the shuttlecock (3).

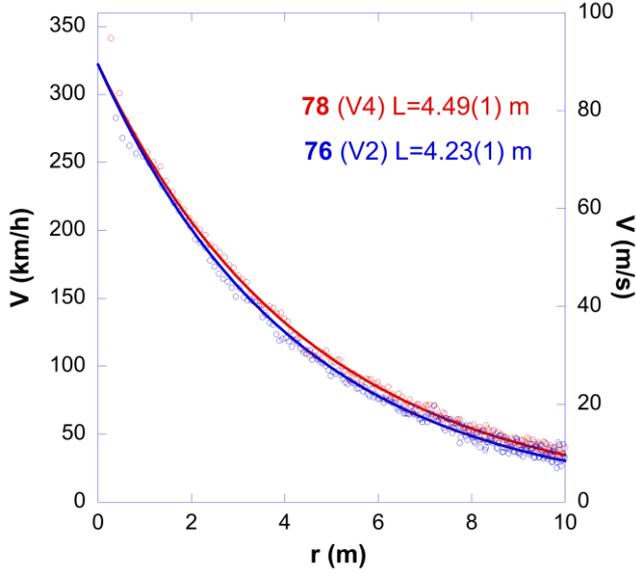

**Figure 6.** Velocity decay after smash shot for Yonex AS20 shuttlecocks. The aerodynamic length is higher for the shuttlecock with fast speed index (78/V4, $L$=4.49(1) m) compared to the one with a slow index (76/V2, $L$=4.23(1) m).

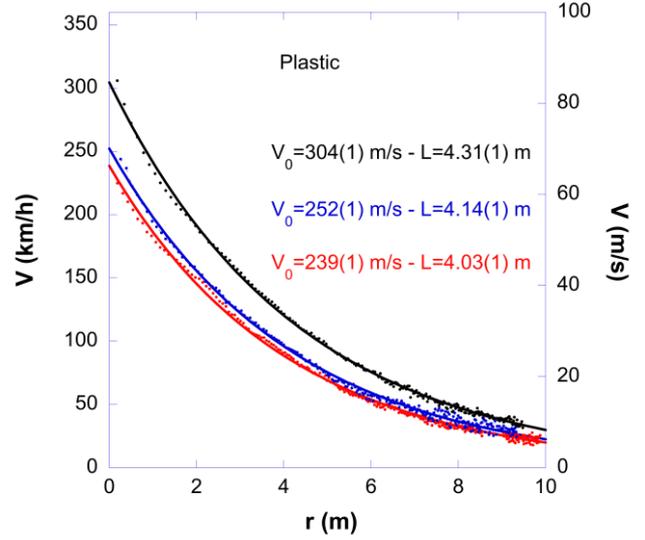

**Figure 7.** Velocity decay after smash shot for plastic Yonex Mavis 300 shuttlecocks for different initial velocities $V_0$. The fit of the $V(r)$ curves with equation (6) shows that the aerodynamic length $L$ increases with $V_0$.

Figure 5 shows the huge acceleration of the shuttlecock during the first 10 cm after the impact with the racket. The velocity increases from 21 to 399 km/h within 1.3 ms, which corresponds to an acceleration of about 8000 g! The measured initial velocity (399 km/h) is slightly higher than the value from the fit at 10 cm from impact (396 km/h), while the extrapolation of the fit at the point of impact gives $V_0$=409 km/h, which much higher than the 21 km/h measured. A precise measurement of $V_0$ requires therefore determining the point of impact with an accuracy of $\approx$ 1 cm, considering that the velocity decays by $\approx$1 km/h every cm.

### 4.3 "Fast" vs "slow" shuttlecocks

How does the change in the speed index of the shuttlecock affects the $V(r)$ curve? Figure 6 shows the velocity decay measured after different smash shots for two Yonex AS20 shuttlecocks with speed index 76/V2 and 78/V4 (table 1). The fit of the experimental curves with equation (6) gives similar initial velocities for both shots ($\approx$320(2) km/h). The aerodynamic length is larger for the 78/V4 ($L$=4.49(1) m) compared to the 76/V2 ($L$=4.23(1) m). Indeed, high-speed shuttlecocks have lower deceleration and higher average speeds than low-speed ones. The shuttles have similar shapes, and their different aerodynamic lengths and decelerations are due to their mass (Table 1), which is reasonably well proportioned to their aerodynamic length.

### 4.4 Plastic shuttlecocks

The main problem with feather shuttlecocks used in competition is that they are easily damaged during play. They therefore need to be replaced frequently to maintain their flight characteristics. During the last decades synthetic shuttlecock emerged, made from plastics, with overwhelming advantages in durability, cost-effectiveness, and consistency in recreational play (15). Plastic shuttlecocks, whose skirts are mainly made of nylon, are becoming increasingly popular in non-competitive badminton, as their flexibility gives them greater durability. However, this flexibility comes at the cost of inferior flight quality compared to feather shuttlecocks. Several studies have shown that the drag coefficient of synthetic shuttlecocks decreases with increasing speed, due to the large deformation of the skirt with the air pressure at high speed (3-5, 10, 13, 26). A recent computational study (29) confirmed how the cross-sectional area of the skirt decreases with an increase in flight speed, leading to a significant reduction in the drag compared to that for an undeformed shuttlecock. Beyond a certain speed, the deformation of the skirt assumes a non-axisymmetric shape with a significant increase in its rate of deformation with speed.

The terminal velocity of the plastic shuttlecock Mavis 300 is lower than the one of the feathers (table 1). Therefore, a shorter aerodynamic length would be expected compared to the feather AS20 shuttles. Figure 7 shows its velocity decay for different initial velocities $V_0$. The fits of the $V(r)$ curves with equation (6) show that the aerodynamic lenght increases with initial velocity, from $L$=4.03(1) m for $V_0$=239 km/h to $L$=4.31(1) m for $V_0$=304 km/h. This confirms that the drag decreases with increasing initial speed. The video 3 and figure 8 show the deformation of the plastic shuttlecock from an initial shape with 16-fold symmetry to square or triangular shape after smash, due to the air pressure on the flexible skirt. The resulting change in the cross-section and drag coefficient of the shuttlecock lower the drag force. As the shuttlecock slows down, the air pressure on the skirt decreases and the initial shape is almost recovered within ≈3 meters, but the radius of the shuttle, and therefore the drag, still depend on speed (3-5, 10, 13, 26). These data, measured directly on a badminton court under real conditions, confirm previous experimental and theoretical analyses concerning the variation in drag coefficient as a function of initial velocity due to the deformation of the skirt of plastic shuttlecocks.

### 5. Effect of slice and reverse slice on velocity decay

The slice shot is another tactical technique, known for causing the shuttle to spin and altering the angle of return. Kitta conducted experiments on feather shuttlecock (1) and concluded that the shuttlecock with spin experiences a marginally larger drag compared to shuttle without spin. A recent study explained why the slice shot of the left-handed player (LH) is more efficient than the slice shot of the right-handed player (RH) (25), due to the chiral nature of a shuttlecock (figure 1). The helical arrangement of the feathers is responsible for its natural counter-clockwise rotation as it propagates through the air. Video 4 shows that the slice shots performed by right-handers induce a natural counter-clockwise spinning. On the contrary, the slice shots performed by left-handers induce an opposite clockwise spinning: the air flow on the feathers then stops the rotation and the shuttlecock, which finally spins naturally counter-clockwise. During this clockwise to counter-clockwise spining for the slice shot of the LH, some kinetic energy is therefore transferred to rotation energy, which slows down more the shuttlecock. Reversively, the reverse slice shots performed by LH induce a natural counter-clockwise spinning, while the reverse slice shots performed by RH induce a clockwise to counter-clockwise spinning.

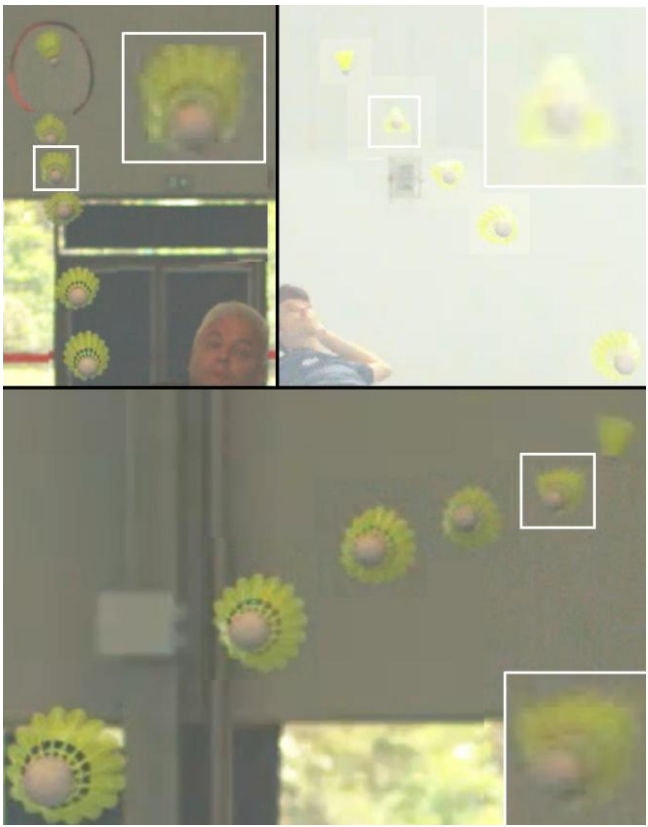

**Figure 8.** Deformation of the skirt of a plastic shuttlecock at high speed after impact with racket, with triangular or square shapes depending on initial velocity. As the shuttlecock slows down and spins, the initial shape is recovered within ≈3 m.

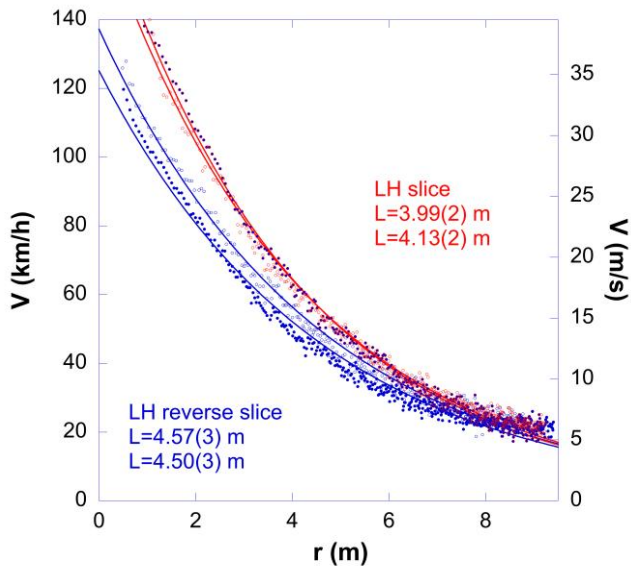

**Figure 9.** Speed decay of a AS20 V3/77 shuttlecock. The aerodynamic length is lower for the for a left-hander (LH) slice (<$L$>≈4.06 m) compared to the LH reverse slice (<$L$>≈4.53 m).

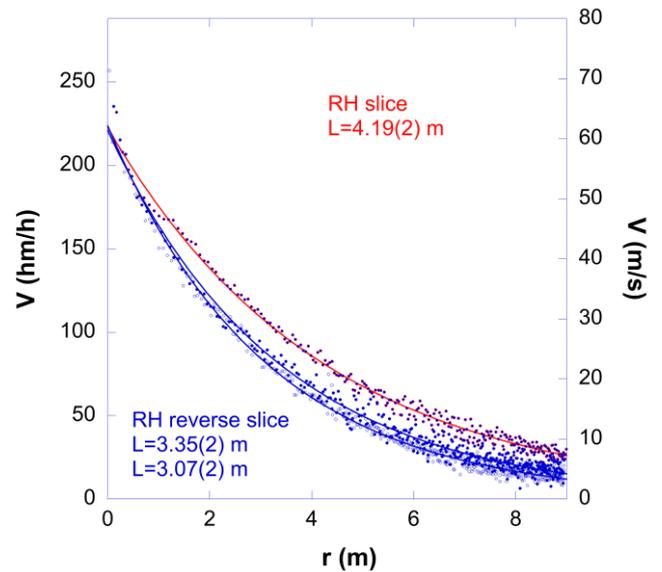

**Figure 10.** Speed decay of a AS20 76/V2 shuttlecock. The aerodynamic length is lower for the right-hander RH reverse slice (<$L$>≈3.21(2) m) compared to the RH slice (<$L$>≈4.19(2) m).

Figure 9 compares the decrease in shuttlecock velocity after slice shots and reverse slice shots for a left-handed player. The $V(r)$ curves show that the speed decreases more rapidly for the left-handed slice shot than for the left-handed reverse slice shot, as characterized by the fit of the curves to equation (6), which gives a shorter aerodynamic length $<L>\approx 4.06$ m for the LH slice compared to $<L>\approx 4.53$ m for LH reverse slice shot. Figure 10 shows that the situation is reversed for RH players (with shots performed with an AS20 76/V2 shuttlecock): the fit gives a longer aerodynamic length ($L \approx 4.19$ m) for the RH slice shot than for RH reverse slice ($<L>\approx 3.21$ m). Figures 9 and 10 show that, compared to a natural counter-clockwise rotation induced by a RH slice or a reverse LH slice, the opposite clockwise rotation induced by a LH slice or a reverse RH slice results in a more effective slowing down of the shuttlecock, which is consistent with a recent symmetry analysis (25).

## 6. Conclusion and lessons learned

This study shows that it is possible to describe with a good accuracy the velocity change of a badminton shuttlecock after smash or slice shots as an exponential decay with distance. The aerodynamic properties of the shuttlecocks, described through their speed index, allows adjusting the shuttlecock to the local atmospheric conditions. The speed decay is very fast as velocity is halved every ≈3.35 m, with important consequences in the game, which are of great interest for players or coaches.

Figure 2 shows that the velocity decays is important within the first meter. A smash at 500 km/h from the backline is equivalent to a smash at ≈400 km/h shot one meter in front of the backline. It is therefore of tactical importance for a player in a defensive position, who has to lift the shuttlecock and who will have to face the smash of the oponent, to push the oponent at the back of the court.

This point is evenmore interesting in terms of time-of-flight of the shuttlecock, which is found by intergrating equation (6):
$$t(r) = \frac{L}{V_0}\left(\exp\left(\frac{r}{L}\right) - 1\right) \qquad (7)$$

Figure 11 shows the measured dependence of the time-of-flight with distance $t(r)$ for a smash performed with a AS20 78/V4. The fit of the data with equation (7) gives $V_0=300(1)$ km/h for this shot, and the aerodynamic length $L=4.48(2)$ m is again in good agreement with the result for another shot in figure 6. Figure 11 shows that the time of flight for an initial speed of 300 km/h increases from ≈440 ms to travel over 10 m to ≈660 ms to travel over 11.5 m. This also highlights the importance of pushing the smasher towards the back of the court. Indeed, for a given initial velocity $V_0$, increasing the distance travelled from 10 to 11.5 metres increases the reaction time to defend the smash by 50%.

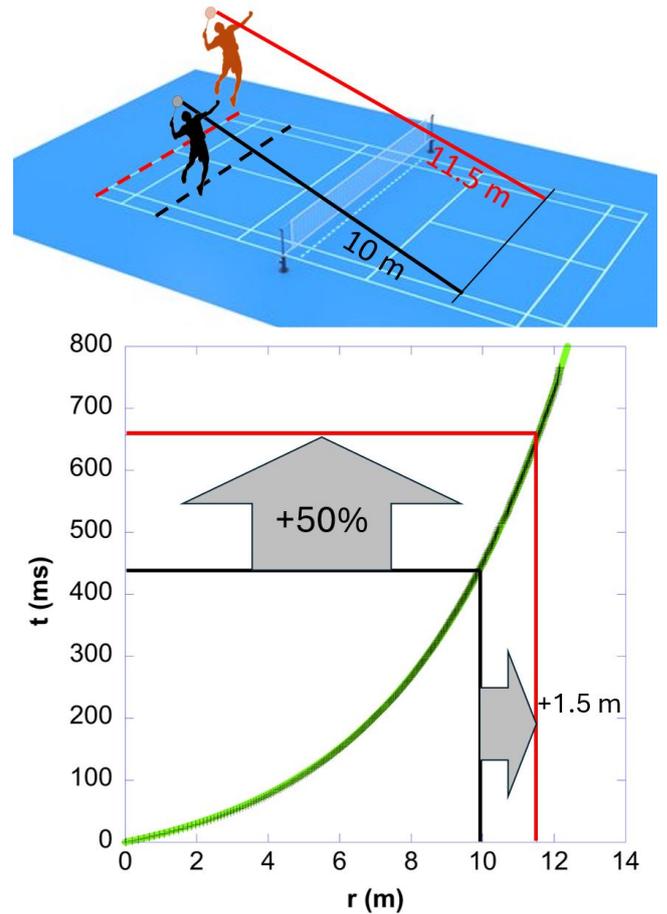

**Figure 11.** Time-of-flight of a AS20 78/V4 shuttlecock after a smash shot with an initial velocity of ≈300 km/h. The time-of-flight increases by ≈+50% for a smash of 11.5 m *vs* 10 m.

For sustainability reasons, the BWF initiative to transition from traditional feather shuttlecocks to synthetic-feather shuttlecocks has introduced equipment changes that could impact gameplay. The flight characteristic of plastic shuttlecocks with an aerodynamic length depending on the initial velocity, makes it difficult to control their trajectory. For this reason, plastic shuttlecock, which have inetresting longevity, cannot fully replicate the precise flight behavior of a high-quality natural feather shuttle. As new synthetic shuttlecocks made of carbon appeared recently on the market (30), it will be interesting to compare the flight characteristics and speed decay of this new generation of shuttlecocks with traditional feathered shuttles, which are currently the benchmark for flight quality.

## Acknowledgements

E.C. would like to thank B. Kersaudy, A. Lièvre, E. Gautier and C. Allaire, the players who participated in this study, and M. Dureault, K. Nakamura and A. Gauthier for technical support. The author also thanks the BWF, the FFBad and especially C. Grandamas for providing acces to the court during IFB 2025 in Rennes. E.C. would also like to thank Gillian Clark and Steen Pedersen for fruitful discussions.